\documentclass[prb,reprint,showpacs,floatfix,epsf,epsfig,onecolumn]{revtex4} 
\usepackage{epsf}
\usepackage{epsfig}
\usepackage{epstopdf}
\usepackage{graphicx}

\begin{document}
\title{Ab initio Studies on Electronic and Magnetic Properties of X$_{2}$PtGa (X = Cr, Mn, Fe, Co) Heusler Alloys}
\author{Tufan Roy$^{1}$, Aparna Chakrabarti$^{1,2}$\footnote{Electronic mail: aparnachakrabarti@gmail.com}}
\affiliation{$^{1}$ Homi Bhaba National Institute, Training School Complex, Anushakti Nagar, Mumbai-400094, India}
\affiliation{$^{2}$ Indus Synchrotrons Utilization Division,Raja Ramanna Centre for Advanced Technology, Indore-452013, India}

\begin{abstract} 
Using first-principles calculations based on density functional 
theory, we probe the electronic and magnetic   
properties of X$_{2}$PtGa (X being Cr, Mn, Fe, Co) Heusler 
alloys. Our calculations 
predict that all these systems possess inverse Heusler alloy 
structure in their respective ground states. Application of 
tetragonal distortion leads to lowering of energy with respect to the 
cubic phase for all the materials. The equilibrium volumes of both 
the phases are nearly the same. These results of our calculations 
indicate that all 
these materials are prone to undergo martensite transition, as has 
been recently shown theoretically for Mn$_{2}$PtGa in the literature. 
Ground state with a tetragonal symmetry of these materials 
is supported by the observation of soft 
tetragonal shear constants in their cubic phase. 
By comparing the energies of various types of magnetic configurations 
of these alloys we predict that Cr$_{2}$PtGa and Mn$_{2}$PtGa possess 
ferrimagnetic configuration whereas Fe$_{2}$PtGa and Co$_{2}$PtGa 
possess ferromagneic configuration in their respective ground states.
\end{abstract}


\maketitle

\section{Introduction}  

Full Heusler alloys (FHA) have drawn considerable attention of the 
researchers as some of them show shape memory alloy property. In 
this family, Ni$_{2}$MnGa is the most studied prototype 
ferromagnetic shape memory alloy. It is reported that Ni$_{2}$MnGa 
shows large magnetic field induced strain (MFIS) as well as 
magnetoresistance effect (MRE), which makes Ni$_{2}$MnGa a potential 
material to be used as sensors, actuators 
etc.\cite{phil-web-1984,apl-sozinov-2002,apl-cbiswas-2005} 
However, the basic drawbacks of this material in terms of 
technological application are its brittleness and the low martensite 
transition temperature (T$_{M}$ which is 210 K)\cite{JPCM-brown-1999}, 
because from the application point of view, it is mandatory that 
T$_{M}$ should be above room temperature. In the literature, it is 
reported that, for this type of FHA systems, the T$_{M}$, Curie 
temperature (T$_{C}$) as well as the inherent crystalline 
brittleness (ICB) are highly dependent on the composition. Hence, 
there is a vast amount of work searching for new Heusler alloys with 
improved properties, which include low ICB, as well as T$_{M}$ and 
T$_{C}$ above room teperature. This is important from the point of 
view of technological application, for example, as actuators, sensors 
etc.\cite{prb-sroy-2009, prb-achakrabarti-2005,prb-barman-2008,apl-achakrabarti-2009,jpcm-khan-2004,apl-mario-2011,apl-stadler-2006,prb-achakrabarti-2013,jalcom-troy-2015,jmmm-troy-2016}
There are also some shape memory alloys found in the literature 
which are termed as high temperature shape memory alloys (HTSMA), 
whose martensite transition temperature is above about 120$^{\circ}$C 
(i.e. 393 K) and they find their application in engines of 
automobiles, turbines, airplanes 
etc.\cite{intmet-otsuka-1999,mettrans-smialek-1973,intmet-yang-1994,mat-tech-99}

There is another class of FHAs which is seen to be half 
metallic in nature.\cite{prl-Groot-1983} In this kind of Heusler 
alloys, density of states of either the majority or the minority 
spin vanishes at the Fermi level and these materials have potential 
application as spintronic devices. Generally, many of the Co-based 
Heusler alloys exhibit this behaviour, like, Co$_{2}$CrGa, 
Co$_{2}$MnGa, Co$_{2}$MnSn.\cite{JPD40HCK,PRB-76-024414-2007} 
However, some metallic Co-based materials,
 like, Co$_{2}$NbSn, Co$_{2}$NiGa are found to exhibit shape memory 
alloy property.\cite{JPSJ58SF,tufanPRB2016,prb-mario-2010,actamater-arroyave-2010,apl-Dai-2005,scrmat-Dadda-2006}
 Hence, it may be worth probing if there are Co-based Heusler alloys 
which show both the properties i.e. shape memory property as well as 
high spin polarization at the Fermi level. One such material, 
Co$_{2}$MoGa, has recently been predicted by us.\cite{tufanPRB2016}

Currently, Co$_{2}$NiGa and related alloys have gained interest among 
the researchers. Some of these have been studied in detail 
theoretically\cite{prb-mario-2010, actamater-arroyave-2010} as well as
have already been prepared experimentally.\cite{apl-Dai-2005, scrmat-Dadda-2006} Interestingly, Co$_{2}$NiGa shows the shape memory alloy 
property as well as 
it possesses different (inverse Heusler alloy) structure compared to 
many of the Co-based systems, which are known to exhibit 
half-metallicity and 
have the conventional Heusler alloy structure in their 
ground state. It is to be further noted that, the Co-based materials 
possess reasonably high Curie temperature. All these above-mentioned
findings in the literature have motivated us to probe the 
Co$_{2}$NiGa-derived Heusler alloy systems. 
It is to be further noted that, in a very 
recent work\cite{jalcom-troy-2015} it has been shown that, 
substitution of Ni by Pt in case of Ni$_{2}$MnGa reduces the 
inherent crystalline brittleness as well as it makes the tetragonal 
state more stable compared to its cubic phase. Besides that, Mario 
et al have observed an enhancement of T$_{C}$ in case of Co doping 
in the Pt based systems.\cite{apl-mario-2011}  So, studying the 
literature\cite{jalcom-troy-2015,apl-mario-2011,equib-bdutta-2014}, 
we can expect  
that, the effect of replacement of Ni by Pt on various physical  
 properties of Co$_{2}$NiGa may turn 
out to be interesting both from fundamental as well as application 
points of view. 

Furthermore, number of valence electrons of the system has been 
shown to play an important role in determining the properties of 
these alloy systems. In this work, therefore, we have discussed 
the changes in the  magnetic, electronic as well as mechanical 
properties of X$_{2}$PtGa as X is varied and the number of valence 
electrons of the X element, which is a first row transition metal 
atom (TM), changes systematically (X being Cr, Mn, 
Fe, Co). Among these materials Mn$_{2}$PtGa is already experimentally 
prepared and theoretically studied.\cite{nature-anayak-2015,prl-anayak-2013,jap-anayak-2015,PRB-wollman-2015} It shows an inverse Heusler
alloy structure and possesses a ferrimagnetic
 configuration as is the case for Mn$_{2}$NiGa.\cite{sunil-jpcm-2014}
We compare our calculated results with the data in the literature, 
wherever available. In what follows, first, we give a brief account 
of the method we used and then we present the results and discussion. 
In the end, the results of this work are summarized and conclusions 
are drawn.

\section{Method}  

The geometry optimization of all the materials have been carried out 
using Vienna Ab initio Simulation Package 
(VASP)\cite{prb-kreese-1996} where the projector augmented wave 
method (PAW)\cite{prb-blochl-1994} is implemented. For the 
exchange-correlation functional, generalized gradient approximation 
has been used over local density 
approximation.\cite{prb-pbe-1996} 
We have used the energy cut-off for plane wave expansion of 
500\ eV. The calculations of energies have been performed with 
a $k$ mesh of 15$\times$15$\times$15 for the 
cubic case and for the tetragonal phase also we have used similar 
kind of $k$ mesh. All the calculations have been carried out with 
the energy and the force tolerance 
were 10 $\mu$eV and 10 meV/\AA, respectively.
For cubic crystals there are three independent elastic constants 
namely $C_{11}$, $C_{12}$ 
 and $C_{44}$. To find these three independent constants we have 
applied three different kind of strains on the geometrically 
optimized cubic structure of the respective systems using VASP code. 
These strains are ($\delta$, $\delta$, $\delta$, 0, 0, 0)($e_{1}$),
 (0, 0, $\delta$$^{2}$/(1- $\delta$$^{2}$), 0, $\delta$)($e_{2}$) 
and ($\delta$, $\delta$, (1+$\delta$)$^{-2}$-1, 0, 0, 0)($e_{3}$).
The applied strain should be as small as possible; in our case we 
have applied strain varying from 0.02 to -0.02 in steps of 0.005.

 It is well known that, all-electron calculations are more reliable 
for the prediction of magnetic properties particularly for the 
systems containing first row transition elements. Hence, we have 
carried out all electron spin polarized relativistic calculations 
on the optimized geometry for all the systems using full potential 
linearised augmented plane wave method 
(FPLAPW).\cite{wien-pblaha-2002} 
For exchange-correlation functional, generalized gradient 
approximation (GGA)\cite{prb-pbe-1996} over local density 
approximation is used. An energy cut-off for plane wave expansion 
was used of about 16 Ry ($R_{MT}$$K_{max}$ = 9.5). The cut-off for 
charge density was G$_{max}$=14. The number of $k$ points in the 
self-consistent-field calculation is 8000 (256)  in the cubic case, 
whereas for the tetragonal case the number is 8000 (635) in the 
reducible (irreducible) Brillouin zone (BZ).
To get detailed insight about the magnetic interaction between the 
constituent atoms, we have used spin-polarized-relativistic 
Korringa-Kohn-Rostoker method (SPRKKR) as implemented in SPRKKR 
package\cite{rep-ebert-2011}, which provides the Heisenberg exchange 
coupling constants (J$_{ij}$). From J$_{ij}$ we have calculated 
T$_{C}$ following the approach of Liechtenstein 
et al.\cite{Liechtenstein-jmmm-1987} The SCF calculations have been 
carried out with the local density approximation as the 
exchange-correlation potential and a $k$ mesh of 
21$\times$21$\times$21 in the BZ were used and further
an angular expansion upto $lmax=3$ was considered for each atom.

\section{Results and Discussion} 

\subsection{Geometry Optimization, Electronic Stability and Possibility of Martensite Transition}

{\it Geometry Optimization} - 
FHAs in the cubic phases crystallize in two types of 
crystal structures, namely, conventional Heusler alloy structure and 
inverse Heusler alloy structure. In case of FHA there are four fcc 
sublattices centered at (0.25, 0.25, 0.25), (0.75, 0.75, 0.75), 
(0.50, 0.50, 0.50), (0.00, 0.00, 0.00), which we label as A, B, C 
and D sublattices, respectively. In case of a conventional Heusler 
alloy structure (with a formula X$_{2}$YZ), the X atom occupies the 
A and B sublattices, the Y atom occupies the C sublattice and D 
sublattice is occupied by the Z atom. This X$_{2}$YZ structure
shows a Fm$\bar{3}$m (number 225) space group. On the other hand, 
in case of inverse Heusler alloy structure (with a formula XYXZ), X 
atom occupies the A and C sublattices and they are termed as X2 and 
X1, respectively, Y atom occupies the B sublattice whereas the
Z atom occupies the D sublattice. This XYXZ structure assumes a 
F$\bar{4}$3m (number 216) space group. Full geometry optimization 
has been carried out for both the possibilities for all the materials.

\begin{table*}
Table~1. Comparison of formation energies between inverse and conventional Heusler alloy structure with different magnetic configurations. Lattice parameters after geometry optimization are also reported.\footnote{Comparison with experiments or previous calculations, wherever data are available \\
$^{b}$Ref.\onlinecite{PRB-wollman-2015}}
\begin{tabular}{|c|c|c|c|c|c|}
\hline Material & Crystal structure & Magnetic Configuration & Formation Energy (kJ/mol) & a$_{init}$(\AA )& a$_{opt}$ (\AA ) \\
\hline&Inverse&FM&-62.63(converged to FIM)&6.10&6.10\\
&Inverse&FIM&\textbf{-62.63}&6.10&6.10\\
{Cr$_{2}$PtGa}&Inverse&NM&+2.88&6.10&5.99\\
&Conventional&FM&+6.36&6.10&6.22\\
&Conventional&NM&+68.92&6.10&6.03\\
\hline
 &Inverse&FM&-72.75&6.13&6.15\\
&Inverse&FIM&\textbf{-117.67}&6.13&6.13\\
&&&&&6.13$^{b}$\\
Mn$_{2}$PtGa&Inverse&NM&+25.45&6.13&5.93\\
&Conventional&FM&-71.29&6.10&6.21\\
&Conventional&NM&+45.98&6.10&5.94\\
\hline &Inverse&FM&\textbf{-99.58}&6.00&6.00\\
&Inverse&FIM&-36.98&6.00&6.02\\
Fe$_{2}$PtGa&Inverse&NM&+47.89&6.00&5.91\\
&Conventional&FM&-29.46&6.00&6.02\\
&Conventional&NM&+37.67&6.00&5.91\\
\hline &Inverse&FM&\textbf{-58.83}&5.95&5.95\\
&Inverse&FIM&-23.22&5.95&5.89\\
 Co$_{2}$PtGa&Inverse&NM&-12.72&5.95&5.90\\
&Conventional&FM&-46.57&5.95&5.93\\
&Conventional&NM&-14.35&5.95&5.90\\
\hline
\end{tabular} 
\end{table*}

{\it Electronic Stability and Ground State Magnetic Configuration} - 
For all the systems studied here (X$_{2}$PtGa, X being Cr, Mn, Fe, 
Co), the formation energies in the cubic structure have been 
calculated and compared (Table 1) for both inverse and conventional 
Heusler alloy structure with different magnetic configurations to 
probe the ground state of these systems both in terms of crystal 
structure and the magnetic configuration. We have considered three 
different kinds of magnetic configurations. These are with 
long-range ferromagnetic (FM), ferrimagnetic (FIM) and nonmagnetic 
(NM) ordering. For FM configuration, all the moments of the X and Pt
atoms are parallel to each other. Under FIM configuration, there is 
one type of magnetic configuration possible, in case of conventional 
Heusler alloy structure, where X and Pt atoms are anti-parallel to
each other. In this case, FIM type of magnetic 
configuration is found to converge to the FM configuration of the 
respective structure. So, consequently, we have not included any 
results corresponding to this magnetic structure (FIM) in Table 1. 
In case of inverse Heusler structure, under FIM configuration, 
there are actually three types of magnetic configurations possible.
 This is because the X1 and X2 atoms are occupying 
crystallographically inequivalent sites.
  In the first case (FIM-1), the moment of Pt atom is taken to be 
anti-parallel to the X atoms (both X1 and X2 atoms are parallel 
to each other). Another ferrimagnetic configuration (FIM-2) is that
where moments of X1 and X2 are anti-parallel to each other and the 
moment of Pt is parallel to X2. The third ferrimagnetic configuration 
(FIM-3) is that where moments of X1 and X2 are anti-parallel to each 
other again and the moment of Pt is parallel to X1. However, the 
FIM-1 configuration converges to the corresponding FM structure. FIM-2 
and FIM-3 turned out to be energetically same for all the systems. 
As a consequence, for the inverse structure, we are only reporting 
the results of FIM-3 configuration in the table and for the latter 
part of the discussion (expressed as FIM hereafter). In 
Table 1, we also report the initial guess value of lattice parameter 
(a$_{init}$) which has been chosen from the literature in case of 
Mn$_{2}$PtGa. For the rest of the materials probed (i.e. Cr$_{2}$PtGa,
  Fe$_{2}$PtGa and Co$_{2}$PtGa), a$_{init}$ has been chosen 
intuitively. In the same table, the optimized lattice parameter 
(a$_{opt}$) for each material after full geometry optimization 
(i.e. relaxing atom positions, unit cell volume and  shape) has 
also been reported. 
More negative value of formation energy in case of inverse Heusler 
alloy structure, compared to their respective conventional Heusler 
alloy structure, suggest that all the systems possess inverse Heusler 
alloy structure, which is consistent with the valence electron 
rule.\cite{PRL-burch-1974} For Cr$_{2}$PtGa and Mn$_{2}$PtGa, inverse 
Heusler alloy structure with FIM configuration and for Fe$_{2}$PtGa 
and Co$_{2}$PtGa the inverse structure with FM configuration possess 
the lowest formation energy, hence these configurations are 
electronically more stable compared to the other combinations. 
Further calculations are carried out on these structures only.

{\it Possibility of Martensite Transition} - 
Heusler alloys may have the potential to be used as shape memory 
alloy device if they undergo a structural phase transition, namely 
martensite transition, from the (low temperature) non-cubic phase 
to a (high temperature) cubic phase above a certain temperature. 
This transition requires that the non-cubic phase must have lower 
energy compared to the cubic phase. We have applied tetragonal 
distortion on the cubic system to probe whether there is a lowering 
of energy with respect to the cubic phase in their respective ground 
state magnetic configuration. A global minimum has been observed for 
all the materials at $c$/$a$ $>$ 1 ($c$ is the lattice parameter 
along z axis and $a$ is the same along x or y direction) as shown in 
Figure 1. It is also seen that the relative volume change between 
the cubic phase and tetragonal phase is nominal (Table 2). It is 
to be noted here that the magnetic configuration for both the cubic
and tetragonal phase has been found and taken to be the same in our 
calculations.
Now we discuss in detail about how we arrive at the volume change 
between the cubic (austenite) phase and the tetragonal (martensite) 
phase as reported in Table 2. We first consider the fully optimized
lattice parameter (Table 1) and assume both the cubic 
and tetragonal phase have the same volume. Then we vary the
ratio of $c$ and $a$ keeping the volume same as that of the
optimized cubic phase (Table 1). If a system shows lowering of energy
for $c$/$a$ $>$ 1, then we obtain the value of that particular 
$c$/$a$ for which energy of the system is lowest. Next, we 
vary the volume of the tetragonal phase keeping that same $c$/$a$.
 From there we get a variation of energy as a function of
total volume of the tetragonal phase. The volume for which the energy
of the tetragonal phase is minimum we consider this as the equilibrium
volume of the tetragonal phase. Then we calculate the relative change
in volume between the optimized cubic and tetragonal phase, which only 
we report in Table 2. All the materials in both cubic and tetragonal 
phases possess almost same volume. 
The maximum variation is about 2.6\% which is for Mn$_{2}$PtGa.
This conservation of the volume and a lowering of energy as a result
of tetragonal distorion indicate 
that these materials are likely to show martensite 
transition.\cite{prb-barman-2008,prb-Rabe-2001}

\begin{figure}[ht]
\includegraphics[width=8cm, height=8cm]{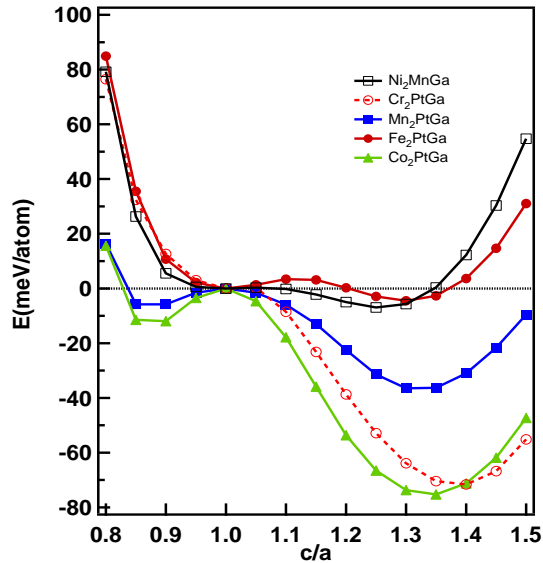}
\caption
{Variation of the total energy of  X$_{2}$PtGa (X = Cr, Mn, Fe, Co) systems in their respective ground state magnetic configurations as a function of $c$/$a$. Equilibrium $c$/$a$ values are expected to be around 1.3 to 1.4 for all the systems. The energy of the tetragonal phase has been normalized with respect to the cubic phase of the respective system. Hence, energy $E$ in the Y-axis signifies the energy difference between the cubic and tetragonal phase. Ni$_{2}$MnGa is presented as a reference material.} 
\label{fig:1}
\end{figure}

It has been observed in the literature that the martensite
transition temperature, $T_{M}$, is approximately proportional 
to $\Delta{E}$ (the energy of the tetragonal phase normalized 
with respect to the cubic phase) and gives reasonable trend of 
the $T_{M}$ across different materials.\cite{prb-barman-2008,AEM14MS} 
Hence we have used the following equation to calculate the $T_{M}$ 
for all these materials. 
\begin{equation}
k_{B}T_{M} = \Delta E
\end{equation}
where \textit{$k_{B}$} is the Boltzman constant and we have used the 
following conversion factor 1 meV = 11.6 K. The calculated $T_{M}$ 
values are listed in Table 2. It is found that except Fe$_{2}$PtGa, 
all the alloys are expected to possess a martensite transition 
temperature which is well above the room temperature and 
Co$_{2}$PtGa yields the maximum value. 

\begin{table}[ht]
Table~2. Calculated lattice parameter $(c/a)_{eq}$ and martensite transition temperature\footnote{Comparison with experiments or previous calculations, wherever data are available \\
$^{b}$Ref.\onlinecite{PRB-wollman-2015}}
\begin{tabular}{|c|c|c|c|c|}
\hline Material&$(c/a)_{eq}$&$\Delta{E}$(meV/atom) &$T_{M}$(K)&$|\Delta{V}|$/V( \%)\\

\hline Cr$_{2}$PtGa&1.41&71.94&834.5&1.00\\
\hline Mn$_{2}$PtGa&1.33&41.06&476.3&2.6\\
&$1.32^{b}$&&&$3.13^{b}$\\
\hline Fe$_{2}$PtGa&1.29&4.40&51.0&0.11\\
\hline Co$_{2}$PtGa&1.37&76.43&886.6&1.50\\
\hline
\end{tabular} 
\end{table}

\subsection{Magnetic and Electronic Property}

{\it Analysis of Magnetic State} - 
In Figure 2 we present the variation of total magnetic moment and 
partial moments as a function of $c$/$a$ for all the systems.
Table 3 presents the 
magnetic moments, percentage spin polarizations ($P_{c}$ and $P_{t}$ 
for cubic and tetragonal phases, respectively) and Curie 
temperatures (calculated from the Heisenberg exchange coupling
parameters as discussed later) of all the magnetic materials in 
their respective cubic
and tetragonal phases. We observe that Cr$_{2}$PtGa and Mn$_{2}$PtGa 
are likely to be ferrimagnetic; on the other hand, 
Co$_{2}$PtGa and Fe$_{2}$PtGa may possess ferromagnetic 
ordering, in their respective ground state. 
The results from the literature, wherever
available, have been presented in the same table 
and it is found that the matching between the data from literature
and our calculations is reasonably good.
 For Cr$_{2}$PtGa, the alignment of spin for Cr atom 
at A sublattice (referred to as Cr2) is opposite to that at C 
sublattice (referred to as Cr1). Same is observed 
in case of Mn$_{2}$PtGa also, resulting in a ferrimagnetic ground 
state as in the case of Mn$_{2}$NiGa.\cite{sunil-jpcm-2014} It is 
worth mentioning that both Cr and Mn atoms possess anti-ferromagnetic 
ground state in their bulk forms. So for this type of (inverse 
Heusler alloy structure) crystal structure,
where X atom is the nearest neighbour of itself, the resulting 
magnetic configuration seems to get influenced by the magnetic 
configuration of the bulk X atom. This is also the case for the other
two materials, namely, Co$_{2}$PtGa and Fe$_{2}$PtGa. 
Both these materials have a long-range ferromagnetic ordering and 
notably both Co and Fe atoms are having ferromagnetic ground state in 
their respective bulk forms.

\begin{table*}[ht]
Table~3. Calculated magnetic moments and Curie temperatures for both austenite ($\mu_{c}$, $T_{C,c}$) and martensite  phases ($\mu_{t}$, $T_{C,t}$). In the parentheses of the second and fifth columns, first two values are corresponding to the inequivalent X atoms (X = Cr, Mn, Fe, Co) and the third value corresponds to the Pt atom. $P_{c}$ and $P_{t}$ are the percentage of spin polarization at Fermi level for cubic and tetragonal phase, respectively. \footnote{Comparison with experiments or previous calculations, wherever data are available \\
$^{b}$Ref.\onlinecite{PRB-wollman-2015}}
\begin{tabular}{|c|c|c|c|c|c|c|}
\hline Material&$\mu_{c}$($\mu_{B}$)&P$_{c}$(\%)&$T_{C,c}$(K) &$\mu_{t}$($\mu_{B}$)&P$_{t}$(\%)&$T_{C,t}$(K)\\
\hline Co$_{2}$PtGa&3.25&80.14&584&2.89&71.59&724\\
&(1.57, 1.60, 0.21)&&&(1.33, 1.58, 0.14)&&\\
\hline Fe$_{2}$PtGa&5.16&1.96&972&5.06&42.34&909\\
&(2.21, 2.86, 0.23)&&&(2.32, 2.75, 0.16)&&\\
\hline Mn$_{2}$PtGa&0.44, $0.44^{b}$&24.18, $23^{b}$ &583, $799^{b}$&0.86, $0.75^{b}$&20.75, $26^{b}$&490, $326^{b}$\\
&(-3.31,3.60, 0.08)&&&(-2.76, 3.42, 0.14)&&\\
\hline Cr$_{2}$PtGa&1.00&74.39&1500&0.28&5.26&1540\\
&(-2.21, 2.95, 0.22)&&&(-2.74, 2.92, 0.04)&&\\
\hline
\end{tabular} 
\end{table*}

\begin{figure}[ht]
\includegraphics[width=8cm, height=8cm]{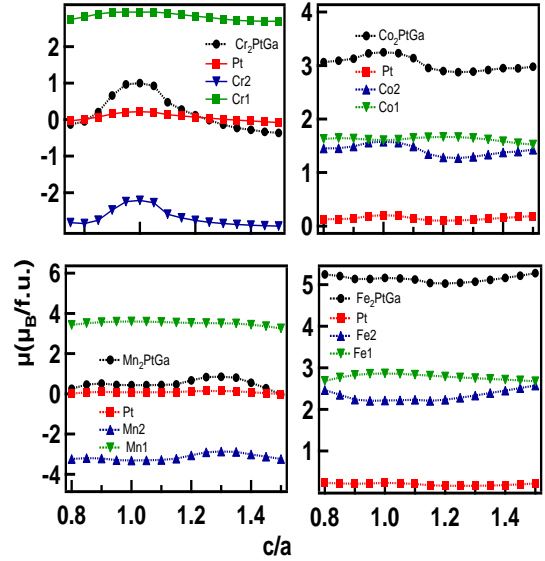}
\caption
{Variation of the total magnetic moment and partial moments of  X$_{2}$PtGa (X = Cr, Mn, Fe, Co) systems as a function of $c$/$a$.} 
\label{fig:2}
\end{figure}

\begin{figure}[ht]
\includegraphics[width=8cm, height=8cm]{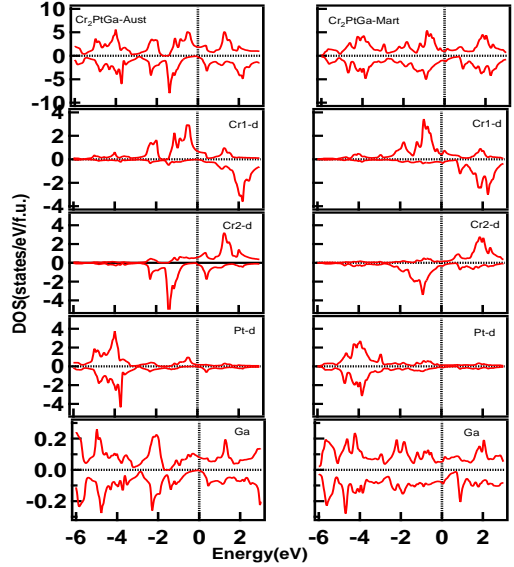}
\caption
{Spin polarized DOS of Cr$_{2}$PtGa (a) Cubic phase (Aust), (b) Tetragonal phase (Mart).}
\label{fig:3}
\end{figure}

\begin{figure}[ht]
\includegraphics[width=8cm, height=8cm]{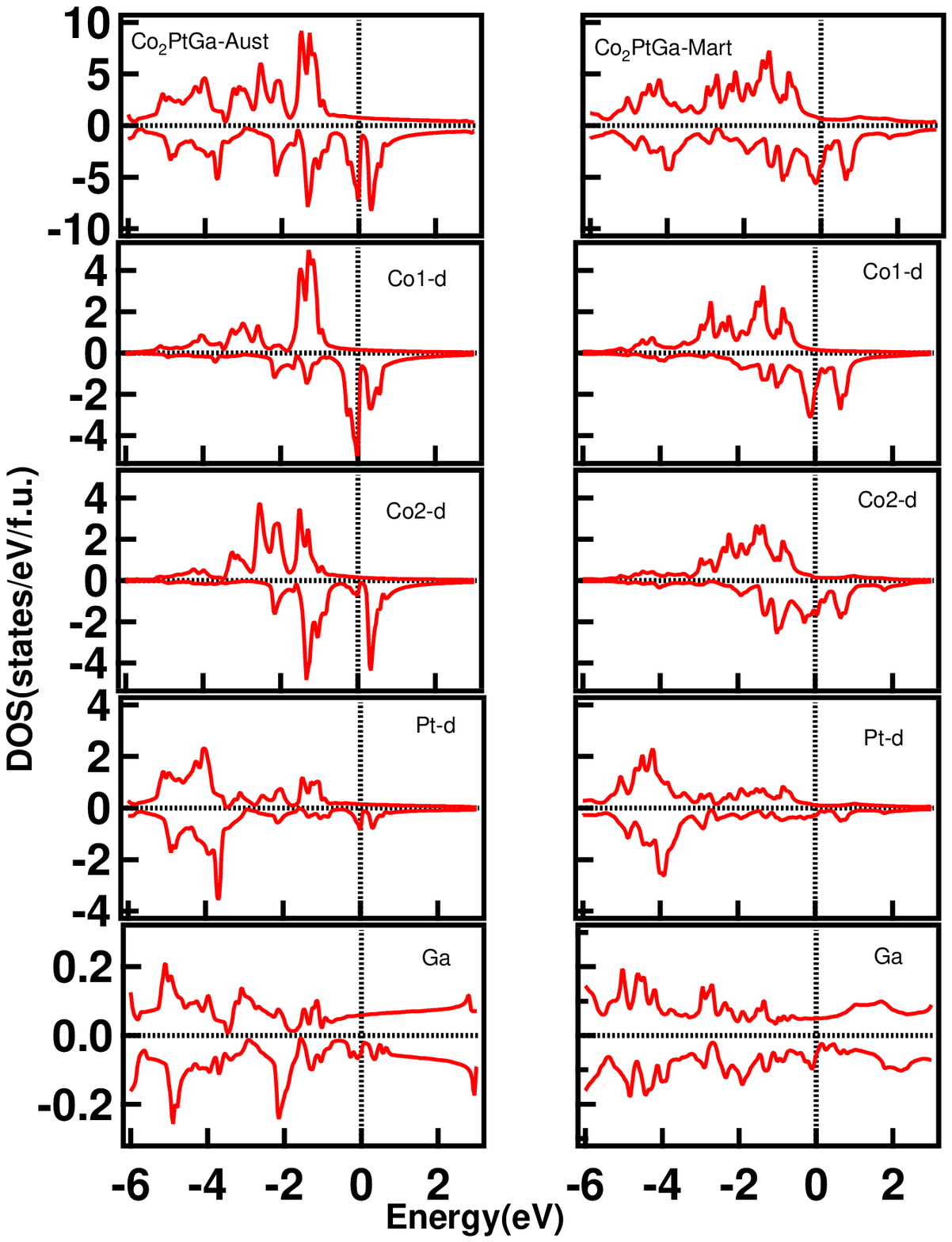}
\caption
{Spin polarized DOS of Co$_{2}$PtGa (a) Cubic phase (Aust), (b) Tetragonal phase (Mart).}
\label{fig:4}
\end{figure}
 
{\it Electronic Density of States to explain Magnetic States} - 
The origin of the ferrimagnetism in both cubic as well as tetragonal 
phase for Cr$_{2}$PtGa can be understood from the electronic density 
of states presented in Figure 3. The moments of Cr1 and Cr2 are 
opposite to each other (Table 3) because of the anti-parallel nature
of the spins of Cr1 and Cr2 atoms below Fermi level, which
is evident from the partial DOS (Figure 3). 
The density of states of Cr1 and Cr2 give rise 
to unequal anti-parallel moments in Cr1 and Cr2 atoms resulting in 
a ferrimagnetic ground state for Cr$_{2}$PtGa. Similar is the case
for Mn$_{2}$PtGa. Figure 4 gives the total and partial DOS for 
Co$_{2}$PtGa in cubic as well as tetragonal phase. It is clearly
seen from the partial DOS of the two Co atoms 
 that the clear exchange splitting 
observed in case of Cr$_{2}$PtGa is not present 
 in case of both the Co atoms of Co$_{2}$PtGa. In both materials
partial DOS of Pt atoms show: (1) the up and down spin DOS are
almost compensated, (2) there is no significant contribution of the 
DOS near the Fermi level. 

\begin{figure}[ht]
\includegraphics[width=8cm, height=8cm]{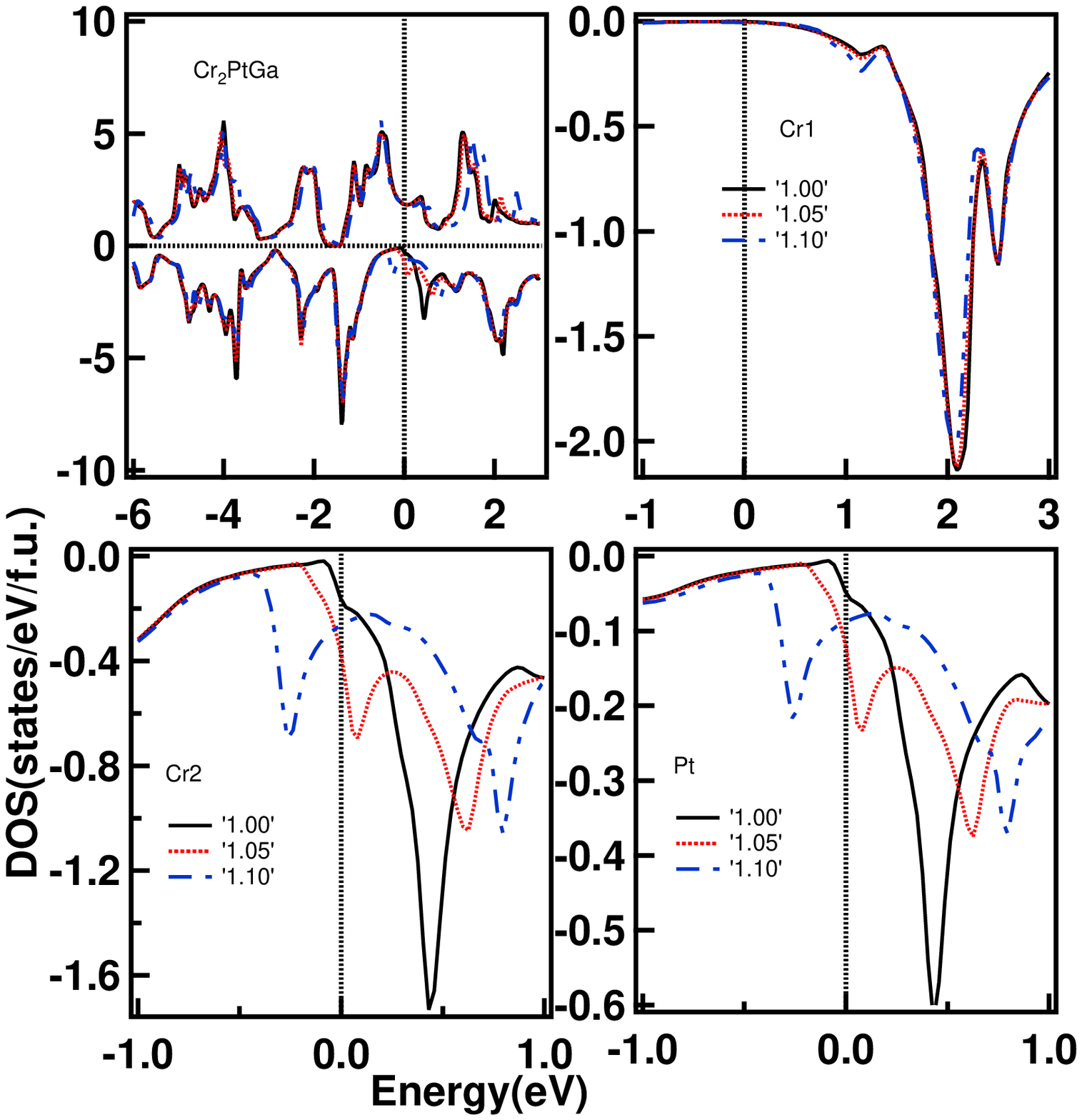}
\caption
{Spin-polarized DOS of Cr$_{2}$PtGa and corresponding minority $3d$ $e_{g}$ level electrons of Cr1 (top panel), as well as Cr2, and Pt atom (bottom panel) as a function of $c$/$a$, maximum value being 1.1.} 
\label{fig:5}
\end{figure}

\begin{figure}[ht]
\includegraphics[width=8cm, height=8cm]{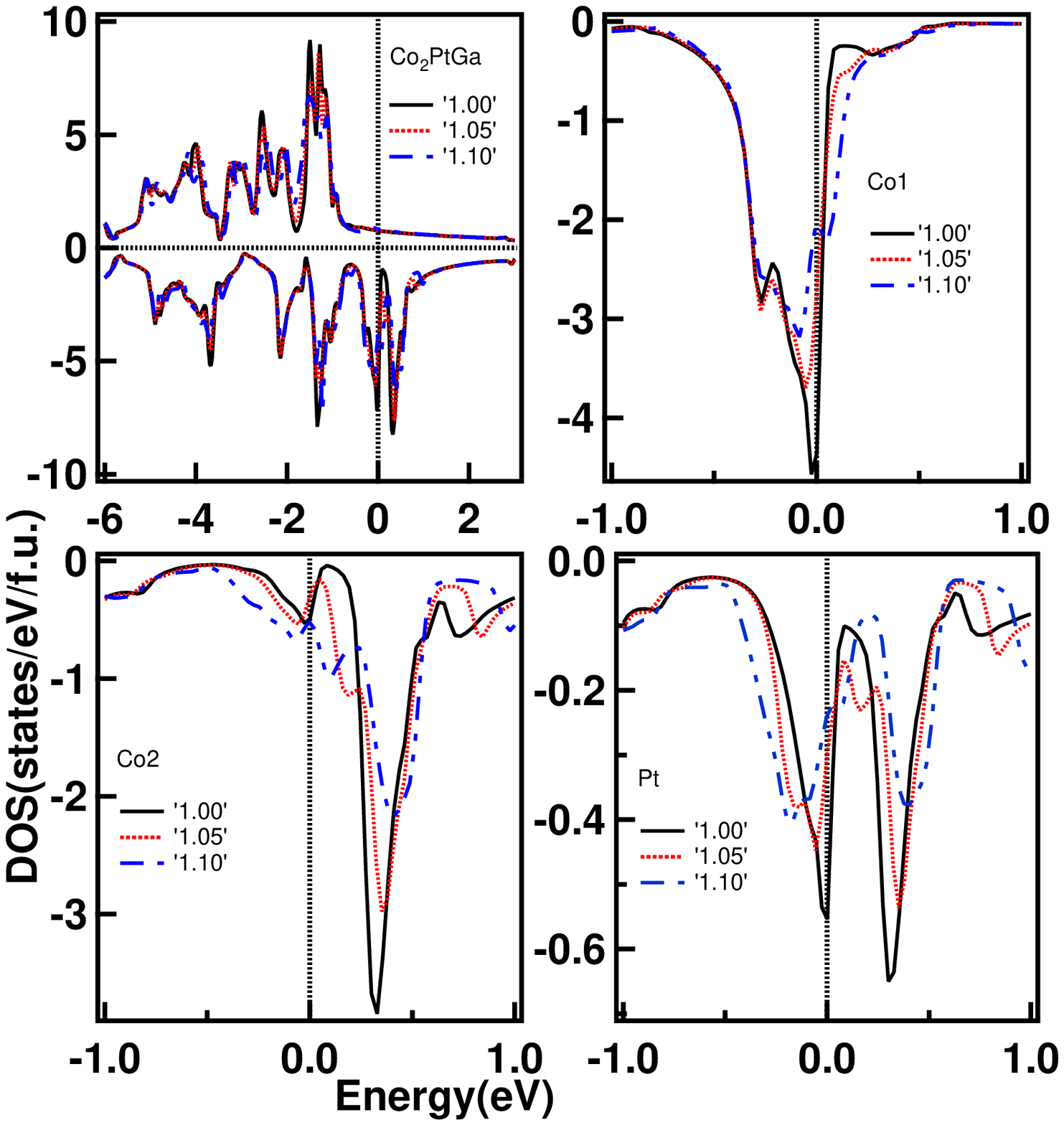}
\caption
{Spin-polarized DOS of Co$_{2}$PtGa and corresponding minority $3d$ $e_{g}$ level electrons of Co1 (top panel), as well as Co2, and Pt atom (bottom panel)  as a function of $c$/$a$, maximum value being 1.1.} 
\label{fig:6}
\end{figure}

{\it Variation of Magnetic Moments as a Function of $c$/$a$} - 
To understand the underlying reason behind the variation of magnetic 
moments as a function of $c$/$a$ (Figure 2) as well as stability 
of the tetragonal phase over cubic
phases we show the spin polarized total density of states along with 
DOS of the minority $d$ ($5d$ for Pt and $3d$ for other TM atoms) 
$e_{g}$ states near the Fermi level for X1, X2 and Pt 
atom as a function of $c$/$a$. We show these results for two typical 
materials, namely, Cr$_{2}$PtGa and Co$_{2}$PtGa in Figure 3 and 4, 
respectively.  The first material is ferrimagnetic and the
other one is ferromagnetic in their respective ground state.

We observe from Figure 2 that there is almost no change in total 
magnetic moment of Fe$_{2}$PtGa as a function of $c$/$a$ which is 
also observed for the cubic and tetragonal phases as is evident
from the values given in Table 3. For Co$_{2}$PtGa we find that, 
there is a maximum at $c$/$a$=1 but in the tetragonal phase the 
total moment decreases, which can be correlated with the increased 
distance between Co1 and Co2 atoms resulting in a weaker long-range 
ferromagnetic interaction between them. On the contrary, for 
Mn$_{2}$PtGa the total moment increases in the tetragonal phase. 
Being the nearest neighbours to each other, Mn1 and Mn2 atoms are 
anti-ferromagnetically coupled. Hence, in this case the increased 
separation between Mn1 and Mn2 makes the anti-ferromagnetic 
interaction weaker in the tetragonal phase compared to its cubic 
phase which results in a larger total moment in the tetragonal phase.
For Cr$_{2}$PtGa the maximum of total magnetic moment at $c$/$a$=1 can
 be correlated with the high spin polarization (75$\%$) at the 
Fermi level in cubic phase. However, in the tetragonal phase the 
spin polarization has decreased significantly (5$\%$) as is also 
seen from the corresponding partial moments on Cr1 and Cr2 atoms. 
This leads to the reduction in the total moment in the tetragonal 
phase with respect to its cubic phase.

{\it $T_{C}$ from Heisenberg Exchange Coupling Constants} - 
We observe that in all the cases, the magnetic moment is found to 
be primarily due to 
the X atoms as is observed from Table 3. A change in the total 
magnetic 
moment between cubic and tetragonal phase is observed. This change 
is maximum for Cr$_{2}$PtGa. 
Because the total moment is substantially lower  
  in the low temperature phase, both Cr$_{2}$PtGa and Co$_{2}$PtGa 
is likely 
to show an inverse magnetocaloric effect.\cite{JPCM-von-2009} 
Next we discuss the Heisenberg exchange coupling paramaters. 
In Heusler alloys, there is a direct exchange interaction and an 
indirect RKKY-type of interaction.\cite{RKKY} The prototype shape 
memory alloy, Ni$_{2}$MnGa, which possesses a conventional Heusler 
alloy structure, a strong ferromagnetic direct interaction exists 
between Ni and Mn, whereas the interaction between two Mn atoms is 
of indirect nature.\cite{PRB28JK,prb-achakrabarti-2013} In case of 
Mn$_{2}$NiGa, with an inverse Heusler alloy structure, there is a 
direct anti-ferromagnetic interaction between the two inequivalent 
Mn atoms which are nearest neighbours to each 
other.\cite{sunil-jpcm-2014} This interaction is much stronger 
compared to the direct exchange interaction observed between Ni
 and Mn in case of Ni$_{2}$MnGa. In Figure 7 we present the Heisenberg
 exchange coupling parameters ($J_{ij}$) between $i$th and $j$th 
atoms for all the systems  X$_{2}$PtGa 
(X = Cr, Mn, Fe, Co) studied here, as a function of inter-atomic 
spacing between them in their cubic phase. The $J_{ij}$'s for intra 
sublattice (X1-X1, X2-X2) and inter sublattice (X1-X2, X1-Pt, X2-Pt) 
are depicted here. It is quite clear from the plots that in all cases 
the X1-X2 interaction is the most dominant one and has a crucial role 
in determining the T$_{C}$ value of the corresponding material. 
A weak oscillatory type of indirect exchange 
interaction is observed for intra sublattice exchange 
interaction X1-X1 and X2-X2, which is an indication of presence 
of RKKY type of interaction in the systems. 

\begin{figure}[ht]
\includegraphics[width=8cm, height=8cm]{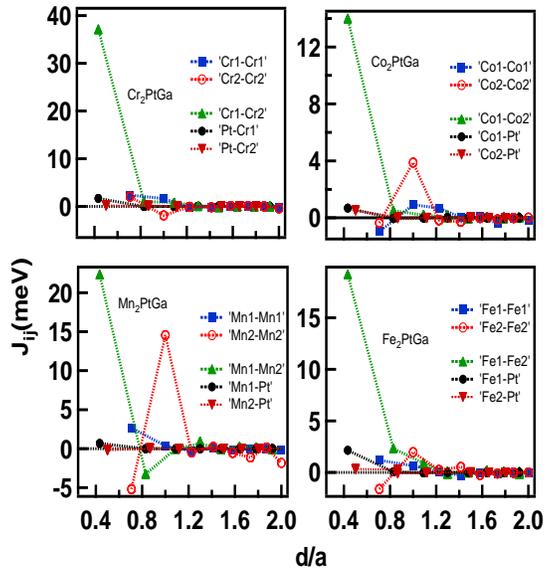}
\caption
{$J_{ij}$ parameters between different atoms of X$_{2}$PtGa 
as a function of distance between the atoms $i$ and
$j$ (normalized with respect to the respective lattice constant).}
\label{fig:7}
\end{figure}

The Curie temperatures for both the austensite and martensite 
phases for all the materials have been calculated from the 
Heisenberg exchange coupling parameters following
 Ref.\onlinecite{Liechtenstein-jmmm-1987}.
In literature it has been mentioned that in the mean field 
approximation, there is an overestimation of Curie 
temperature.\cite{prb-sosioglu-2005,prb-sosioglu-2005-2,prb-rusz-2006}
 It has been argued that in the mean field approximation, spin 
fluctuations are neglected and the magnetic moments are taken to be 
more rigid, which causes an overestimation of Curie temperature.
It is known that a change in $T_{C}$ is observed during a structural 
transition.\cite{PRB-wollman-2015} This has been 
 experimentally observed by 
Khovailo et al\cite{prb-Khovailo-2004} for Ni-Mn-Ga alloy system. 
This change is because of the differences in the partial moment and 
interatomic spacing between the two phases. The strength of 
hybridization between the atoms depends on their separation. 
This hybridization and the moment of the interacting atoms control 
the strength of exchange interaction between the 
atoms\cite{prb-bose-2011}, and consequently, the $T_{C}$ of the 
corresponding system as has been observed from Table 3. The Curie 
temperatures for all the 
systems are observed to be much higher compared to the 
room temperature for the both cubic and tetragonal phases. 
 
{\it Spin Polarization} - 
Table 3 also suggests that the spin polarization ($P$) at Fermi 
level for Co$_{2}$PtGa is the highest in both the cubic and 
tetragonal phases compared to all the other materials. This is much 
higher from the spin polarization of the prototype shape memory 
Heusler alloy, Ni$_{2}$MnGa. This can be understood from Figure 4, 
which presents the spin projected DOS of cubic as well as tetragonal 
phase of this material. We can see that in both the cases, the 
minority spin has a significantly larger contribution to the DOS 
at the Fermi level compared to the majority spin. 

{\it Stabilization of Tetragonal Phase versus the Density of States} - 
The stabilization of the tetragonal phase over the cubic phase of 
the shape memory Heusler alloys has been argued by the band 
Jahn-Teller mechanism.\cite{JPSJ58SF, prb-barman-2005} We have 
recently shown\cite{tufanPRB2016} for a series of Ni and Co-based 
conventional FHAs, i.e. the X$_{2}$YZ type, that the closeness of 
minority DOS peak of the X atom, with $3d$ $e_{g}$ symmetry is 
closely related to the possibility of martensite transition.
Here, in this work, all the materials possess inverse heusler alloy 
structure (XYXZ). A detailed analysis of density of states shows that 
for these systems the stability of the tetragonal phase over its 
cubic phase can be correlated with the presence of minority DOS 
corresponding to the 
$d$ $e_{g}$ peak close to the Fermi level for X2 and Pt atom in 
their respective austenite phase. From Figure 3 we observe that, the 
Cr2 atom's $3d$ $e_{g}$ minority DOS peak is located just above the 
Fermi level (about +0.43 eV). Under tetragonal distortion, this peak 
is split into two. For $c$/$a$=1.10, one part of the peak enters below 
Fermi level and other part moves above the Fermi level. The same kind 
of changes is also observed with the Pt $5d$ $e_{g}$ minority DOS peak 
which is located at the same energy as Cr2 atom in the cubic phase. 
This redistribution of density of states for these two atoms 
leads to the reduction of free energy and to stabilization of
 the tetragonal phase over the cubic phase. But for Cr1 atom, 
there is not much change observed in its minority $3d$ $e_{g}$ 
density of states under tetragonal distortion. 

In case of Co$_{2}$PtGa also, we find the same kind of redistribution 
of minority $3d$ $e_{g}$ DOS of Co2 and Pt atom under the tetragonal 
deformation (Figure 4). The single DOS peak of $3d$ $e_{g}$ minority 
spin of 
Co2 atom is observed at +0.33 eV. Under tetragonal deformation this 
single peak splits into two parts, major part of the peak moves 
into the higher energy side (+0.35 eV for $c$/$a$=1.05) and minor 
part of the peak moves toward the Fermi level (+0.19 eV for 
$c$/$a$=1.05). Here also we find that there is not much change in the 
$3d$ $e_{g}$ DOS of Co1, causing not much of a change in the moment of 
Co1 atom as a function of $c$/$a$.
For both the materials Cr$_{2}$PtGa and Co$_{2}$PtGa, we find that 
 the tetragonal distortion causes the change of peak position of 
$d$ $e_{g}$ DOS primarily of the 
X2 atom and Pt atom. The X2 atom's moment contributes significantly
toward the total moment. This is the reason the 
variation of total moment as a function of $c$/$a$ follows primarily
 the variation of mainly X2 atom’s moment (Figure 2). For all the 
materials studied here, we find mimimum change in the moment of 
X1 as a function of $c$/$a$. 
We note here that Luo et. al.\cite{intmet-luo-2013} have observed 
for Mn$_{2}$NiGe which is also an inverse Heusler alloy, the 
variation of magnetic moment of the two in-equivalent Mn atoms 
follows different trend because of their different chemical 
surroundings leading to different hybridization effects.

{\it Tetragonal Ground State versus Tetragonal Shear Constant} - 
While we show from the electronic properties, the possibility of 
the materials to be prone to undergo tetragonal distortion,  we
 now focus on the mechanical properties, specifically the tetragonal
shear constant ($C^\prime$). The elastic constants of all the 
materials have been calculated in their cubic phase. 
For the cubic crystal the elastic stability criteria are-
$C_{11}$ $>$ 0; $C_{44}$ $>$ 0; $C_{11}$-$C_{12}$ $>$ 0; 
$C_{11}$ + 2$C_{12}$ $>$ 0.\cite{prb-wu-2007} It is observed that 
all the alloys satisfy these conditions but the 3rd condition is 
not satisfied by some of the materials. This leads to a negative 
value of $C^\prime$ for these. Rest of the materials show small 
positive values of it.
This softening of $C^\prime$ indicates the instability of the 
cubic phase. For the prototype shape 
memory alloy, Ni$_{2}$MnGa, the softening of $C^\prime$ is already 
observed from experiment.\cite{prb-Manosa-1997}
 For Cr$_{2}$PtGa and Fe$_{2}$PtGa, $C^\prime$ is a small positive 
quantity, with values 7.56 GPa and 11.80 GPa, respectively. 
On the other hand, Co$_{2}$PtGa and Mn$_{2}$PtGa exhibit negative
values for the same.

\begin{table*}[h]
Table~4. Mechanical properties of the austenite phase of materials
\begin{tabular}{|c|c|c|c|c|c|c|c|c|c|c|}
\hline Material&C$_{11}$&C$_{12}$&C$_{44}$&C$^{\prime}$&B&G$_{V}$&G$_{R}$&G$_{V}$/B&C$^P$&$\Theta$$_{m}$(K)$\pm$ 300\\
&(GPa)&(GPa)&(GPa)&(GPa)&(GPa)&(GPa)&(GPa)&&(GPa)&\\
\hline Cr$_{2}$PtGa&151.08&135.96&99.86&7.56&141.00&62.94&16.97&0.45&36.90&1445\\
\hline Mn$_{2}$PtGa&120.11&140.67&100.32&-10.28&133.82&56.08&-30.36&0.42&40.35&1262\\
\hline Fe$_{2}$PtGa&190.53&166.93&109.87&11.80&174.80&70.64&25.40&0.40&57.06&1679\\
\hline Co$_{2}$PtGa&140.18&199.71&110.22&-29.77&179.87&54.22&-125.11&0.30&84.49&1381\\
\hline
\end{tabular} 
\end{table*}

Cauchy pressure ($C^P$) is defined as the difference between 
$C_{12}$ and $C_{44}$. The value of $C^P$ indicates the nature 
of bonding in a system.\cite{pettifor,prb-mark-2012} For all the 
systems studied here $C^P$ is found to be positive. 
 Pugh phenomenologically related 
the ratio between shear modulus and bulk modulus ($G$/$B$) to the 
inherent crystalline brittleness of a material.\cite{pugh} In the 
estimation of shear modulus we have followed the formalism given 
by Voigt\cite{voigt}($G_{V}$) over given by 
Reuss\cite{reuss}($G_{R}$). This is because $G_{R}$ is largely 
underestimated for the materials which show softening of $C^{\prime}$;
 the reason behind this has been discussed in detail in our previous 
work.\cite{jalcom-troy-2015} If the ratio ($G$/$B$) is greater than 
$\sim$0.57, the corresponding crystalline material is supposed to be 
inherently brittle. According to our calculations, Co$_{2}$PtGa is 
predicted to possess the lowest inherent crystalline brittleness. 
 The upper panel of Figure 8 shows an inverse linear 
relationship between $C^P$ and $G$/$B$ (= $G_{V}$/$B$). This inverse 
linear relationship between these two parameters is already reported in
 literature for a large number of systems.\cite{Niu,jalcom-troy-2015} 
 Figure 8 (a) suggests that the ICB of Co$_{2}$PtGa is quite low 
with respect to that of Ni$_{2}$MnGa. The bottom panel of Figure 8 
shows the variation of 
$G$/$B$ as a function of atomic number (Z) of the X atom of
 X$_{2}$PtGa (X = Cr, Mn, Fe, Co). It is to be noted that 
the value of $G$/$B$ is lowest for Z = 27, i.e. for Co$_{2}$PtGa. 
In literature\cite{acta-Gu-2008} it has been mentioned that the 
materials with lower value of $G$ are likely to undergo shear 
deformation and become more prone to ductility instead of brittle 
fracture 
when strain is applied which leads to a higher inherent crystalline 
brittleness. 

\begin{figure}[ht]
\includegraphics[width=7cm, height=9cm]{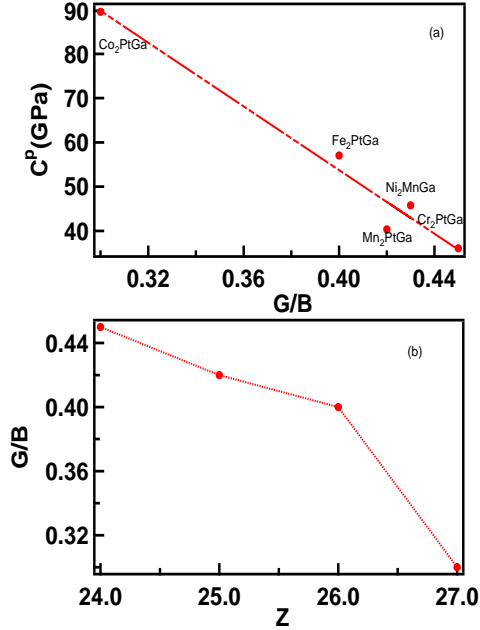}
\caption
 {(a) $C^{P}$ versus $G_{V}$/$B$ plot: A linear fitting has been done, 
 which shows an inverse linear relationship between them. Ni$_{2}$MnGa is 
 used as a reference material. (b) $G$/$B$ versus $Z$ plot. 
 $Z$ is the atomic number of the X atom.}
\label{fig:8}
\end{figure}
 
{\it High Melting Temperature of the Studied Systems} - 
For high temperature application of a material, it is mandatory to 
have very high melting temperatures. Fine et. 
al.\cite{scr-met-fine-1984} have correlated the elastic constant 
$C_{11}$ of the cubic metallic as well
as interametallic systems with the melting temperature by the 
following empirical relationship:
\begin{equation}
\Theta_{m} = 553 K + 5.91\times C_{11}(GPa) \pm 300 K
\end{equation}
Calculated $\Theta$$_{m}$ values for all the systems (Table 4) 
are predicted 
to be sufficiently high for a possible application as SMA even 
at high temperature, specifically for Co$_{2}$PtGa, which is 
predicted to possess $T_{M}$ (around 886 K) well above the room 
temperature.

\section{Conclusion}

From first-principles calculations, we study the  electronic and
magnetic properties of X$_{2}$PtGa (X being Cr, Mn, Fe, Co)
 Heusler alloys. We predict a few new materials with Pt being an 
essential ingredient. By comparing the energies of various types 
of magnetic configurations, we predict that Cr$_{2}$PtGa and 
Mn$_{2}$PtGa possess a ferrimagnetic configuration, whereas, 
Fe$_{2}$PtGa and Co$_{2}$PtGa possess a long-range ferromagneic 
ordering in their respective ground states. Analysing the electronic, 
magnetic and  mechanical properties of all these materials, we 
predict that, Co$_{2}$PtGa, Cr$_{2}$PtGa and Fe$_{2}$PtGa are found 
to be three new full Heusler alloy systems, which are likely to show 
the martensite transition. Among these Co$_{2}$PtGa is likely to 
possess the highest spin polarization at the Fermi level for both 
the cubic and tetragonal phases. It also exhibits the lowest 
inherent crystalline brittleness as well as the 
highest martensite transition ($T_{M}$), melting temperature 
($\Theta_{m}$) and Curie temperature ($T_{C}$); all of these well 
above the room temperature. 

\section{Acknowledgement}
Authors thank P. D. Gupta and P. A. Naik for encouragement throughout 
the work. Authors 
thank S. R. Barman and C. Kamal for useful discussion and Computer 
Centre, RRCAT for technical support. TR thanks HBNI, RRCAT for 
financial support.

\end{document}